\documentclass[12pt]{article}
\def\beq{\begin{equation}}
\def\eq{\end{equation}}
\def\eeq{\end{equation}}

\def\centeron#1#2{{\setbox0=\hbox{#1}\setbox1=\hbox{#2}\ifdim
\wd1>\wd0\kern.5\wd1\kern-.5\wd0\fi
\copy0\kern-.5\wd0\kern-.5\wd1\copy1\ifdim\wd0>\wd1
\kern.5\wd0\kern-.5\wd1\fi}}
\def\ltap{\;\centeron{\raise.35ex\hbox{$<$}}{\lower.65ex\hbox{$\sim$}}\;}
\def\gtap{\;\centeron{\raise.35ex\hbox{$>$}}{\lower.65ex\hbox{$\sim$}}\;}
\def\gsim{\mathrel{\gtap}}
\def\lsim{\mathrel{\ltap}}

\def\CO{{\cal O}}

\begin{document}

\begin{titlepage}

\begin{center}
\vspace*{-1cm}

\hfill SCIPP 07/13  \\
\hfill RU-NHETC-07-10 \\
\vskip .70in
{\LARGE \bf Higgs Physics as a Window } \\
\vskip .1in
%\vspace{.1in}
{\LARGE \bf Beyond the MSSM  }
\vskip .1in
%\vspace{.1in}
{\LARGE \bf(BMSSM) }
\vskip 0.4in
{\large  Michael Dine}$^1$, {\large Nathan Seiberg}$^2$ and {\large Scott Thomas}$^3$

\vskip 0.2in

$^1${ \em Santa Cruz Institute for Particle Physics\\
University of California \\
Santa Cruz, CA 95064}

\vskip 0.15in

$^2${ \em School of Natural Sciences \\
Institute for Advanced Study \\
Princeton, NJ 08540}

\vskip 0.15in

$^3${\em New High Energy Theory Center \\
Department of Physics \\
Rutgers University \\
Piscataway, NJ 08854}

\vskip 0.55in

\end{center}

\baselineskip=16pt

\begin{abstract}

 \noindent
We interpret the current experimental limit on the lightest Higgs
boson mass to suggest that if nature is supersymmetric, there are
additional interactions beyond those of the MSSM coming from new
degrees of freedom around the TeV scale. Within an effective field
theory analysis, the leading order corrections to the MSSM are
described in terms of only two operators. This provides a highly
constrained description of Beyond MSSM (BMSSM) physics. The scalar
Higgs spectrum as well as the chargino and neutralino spectrum and
couplings are modified in a distinctive way. These operators can
be generated by a variety of microscopic mechanisms.

\end{abstract}

\end{titlepage}

\baselineskip=17pt

\newpage

% 2 colum format below

%\pacs{PACS numbers:
%\hfill SU-ITP-03-13 \hfill hep-th/yymmnnn}

%]

% --------------- Introduction -----------------------------

\section{Introduction}

LEPII placed the problem of Higgs physics at the forefront of
supersymmetry phenomenology. While earlier one might have viewed
the Higgs fields as just one of the many features of low energy
supersymmetric models, the constraints on the Higgs mass are now
problematic.  Within the Minimal Supersymmetric Standard Model
(MSSM)\footnote{The phrase MSSM is used here to refer to the
particle content of the model only, not to any features of the
Lagrangian. Similar statements apply later to what is referred to
as the NMSSM.} the Higgs sector occupies a special place.  Unlike
the spectrum of squarks, sleptons and gauginos, which are
determined by many parameters, the Higgs spectrum is quite
constrained. There are only two, currently unknown, tree-level
parameters, and plausible assumptions about fine tuning restrict
these significantly. At the classical (and renormalizable) level,
the mass of the lightest Higgs is lighter than the mass of the $Z$
boson. There are significant corrections to the mass arising from
loops of top quarks and squarks.  But in order that these effects,
by themselves, account for the Higgs mass, the top squarks must be
quite massive -- so massive that the theory appears finely tuned
-- or the top squarks must be highly mixed. This suggests that
{\it if} low energy supersymmetry is important to the solution of
the hierarchy problem, there are likely to be additional degrees
of freedom in the theory beyond those of the MSSM.

There are a number of well-studied candidates for such additional
physics (Beyond the MSSM, or BMSSM physics)
\cite{nmssm, gaugeref, gaugerefb, triplets,mohapatra}.  The addition of a
chiral singlet field gives the Next to Minimal Supersymmetric
Standard Model (NMSSM).
One can contemplate structures with more
chiral fields, gauge interactions, and strongly interacting gauge
theories. This additional physics cannot be too far away; it could
well be at scales near the masses of the Higgs particles. However,
this dynamics might lie at somewhat higher scales (hierarchies of
SUSY breaking scales are familiar, for example, within
gauge-mediated models). If this new dynamics lies at an energy
scale, $M$ above the typical masses of the MSSM fields, one can
organize the analysis by studying an effective Lagrangian from
which physics at scale $M$ has been integrated out.  The effective
action analysis turns out to be {\it extremely} simple, and will
allow us to delineate both phenomenologically interesting regions
of parameter space, as well as robust consequences of such extra
dynamics for the Higgs spectrum.
An effective action analysis
of the MSSM Higgs sector has also been considered in earlier work
\cite{Brignole}.

The power of the effective action follows from the organization of
operators in increasing powers of $1/M$.  For a given observable,
only a small number of operators, with the smallest power of $1/M$
contribute.  Normally, the power of $1/M$ follows from the
dimension of the operator.  However, in theories with approximate
symmetries, the notion that the power of $1/M$ can be larger is
familiar; chirality violating dipole moment operators are perhaps
the most common example.  More generally, for any operator $ \int
d^4x~ (a/M^n) { \cal O}$ (for some constant $a$), the {\it
effective scaling dimension} with respect to the scale $M$ is
$D_{\rm eff}[{\cal O}] = 4 + n$. The same phenomenon arises in
approximately supersymmetric theories. Because supersymmetry
restricts the structure of the operators in the effective theory,
the effective dimension can be larger than the naive dimension. In
particular, we will see below various dimension four operators,
which are suppressed by $1/M$ (or $1/M^2$) -- hence their
effective dimension is five (or six).

This can happen starting with a dimension five
operator which is suppressed by $1/M$, and integrating out
an auxiliary field proportional to a scale, $\mu$ of order
the electroweak scale (such as the
familiar $\mu$ parameter of the MSSM).
In this case the dimensionless coefficient of the dimension four
operator is
 \beq
 |\mu |/M \ll 1.
 \eeq

Supersymmetry breaking operators also generate such corrections.
The MSSM is a renormalizable theory which includes soft
supersymmetry breaking terms. We can add to this theory hard
supersymmetry breaking terms provided they are suppressed by
appropriate powers of $1/M$. Again, such hard supersymmetry
breaking operators have effective dimension larger than their
naive dimension.  Their coefficients are suppressed by
 \beq
 m_{\rm SUSY} /M \ll 1
 \eeq
where $m_{\rm SUSY}$ is a characteristic scale of the
supersymmetry breaking terms in the new physics sector.  We will
see several examples of this phenomenon in what follows.

We will take $m_{\rm SUSY} \sim \mu $ of order a few hundred GeV
and will contemplate $M$ of order a few TeV.  Then we can organize
our computations in a power series in
 \beq
 \epsilon \sim {|\mu| \over M} \sim {m_{\rm SUSY} \over M} \ll 1.
\eeq At leading order in $\epsilon$, we will see that there are only
two operators \cite{strummia}. In terms of the coefficients of these
operators, one can write very simple formulas for the full spectrum.
The operators respect different symmetries, with important
consequences for the qualitative, as well as detailed quantitative,
features of the spectrum.

It is necessary, in the analysis, to distinguish different regions
of
 \beq \eta \equiv {\cot\beta}. \eeq
For moderate $\eta$, the operators at order $\epsilon$ describe
the dominant contribution of BMSSM physics to the Higgs spectrum.
Indeed, the MSSM particle content with these two operators may be
defined as the simplest Beyond Minimal Supersymmetric Standard
Model (BMSSM). For sufficiently small $\eta$, we have a double
expansion in $\eta$ and in $\epsilon$. Depending on the relative
size of $\epsilon$ and $\eta$, a different subset of terms of
order $\eta^2,\ \eta \epsilon$ and $\epsilon^2$ are the dominant
ones.

These operators can have important effects on the phenomenology of
the Higgs sector. Interestingly, the lowest order
operators are not at present bounded
by, for example, precision electroweak measurements. For quite
plausible values of couplings and scales, they can raise the mass
of the lightest Higgs particle appreciably, and alter relations
among the Higgs masses.  Features of the Higgs sector can thus
serve as probes for surprisingly high energy phenomena.

In the effective Lagrangian analysis, there are two types of
questions one can naturally ask about the Higgs sector:  what
measurements would establish that additional degrees of freedom
beyond those of the MSSM are present, and how can one characterize
deviations from MSSM predictions. As we will explain in the next
section, it is possible that the MSSM cannot explain the value of
the lightest Higgs mass.  Measurement of the stop masses could in
fact rule out the MSSM; unless the stops are sufficiently heavy or
very highly mixed, the Higgs mass computed within the minimal
model could simply be too small.  Similarly, as we will discuss,
the MSSM leads to certain relations among the various Higgs
masses. If these relations are not experimentally satisfied, the
MSSM will have to be extended to the BMSSM. In this case,
precision measurements of the Higgs spectrum could thus
determine the coefficients
of the new operators.

In the next section we review some well known facts about the
Higgs sector in the MSSM.  In section 3 we introduce the general
operator analysis, exhibit the small set of operators at low
dimension which can contribute to the Higgs potential, and discuss
their properties.  We consider their effects on the various Higgs
masses and how measurements of the spectrum can establish whether
or not additional interactions beyond those of the MSSM are
required.  In section 4 we turn to more microscopic models,
discussing how these operators arise in the NMSSM, models with
triplets of chiral fields, and theories with extended gauge
interactions, establishing plausible values of the coefficients.
In these models, we discuss ranges of parameters for which the
operator analysis is applicable.  In the conclusions, we discuss
the prospects for connections with experiment, mentioning the
little hierarchy problem and possible scales of new physics.
Finally, in the Appendix the (B)MSSM is compared with the most
general two Higgs doublet model and various symmetries which are
present for certain values of the parameters are discussed.

% ----------------------------------------------------

\section{Review of the MSSM Higgs Sector}

\subsection{Tree level}

The tree-level MSSM Higgs potential with fields $H_u$ and $H_d$
receives contributions from the Higgs soft masses, the
superpotential Higgs mass $\mu$-term $\int d^2\theta \mu H_u H_d$,
and $SU(2)_L \times U(1)_Y$ $D$-term quartic interactions
$$
V = \widetilde m_{H_u}^2 H_u^{\dagger}H_u + \widetilde m_{H_d}^2
H_d^{\dagger}H_d
     - \left( m_{ud}^2 H_u H_d + ~{\rm h.c.} ~\right)
     ~~~~~~~~~~~~~~~~~~~~~
$$
\beq ~~~~~~~~+ {g^2 \over 8} \left[
  (H_u^{\dagger}H_u +  H_d^{\dagger}H_d )^2
   - 4 (H_u H_d)^{\dagger} ( H_u H_d ) \right]
+ {g^{\prime 2} \over 8} (
  H_u^{\dagger}H_u -  H_d^{\dagger}H_d )^2
  \label{VMSSMtree}
\eq where
$$
\widetilde m_{H_u}^2 = |\mu|^2 + m_{H_u}^2
$$
\beq
\widetilde m_{H_d}^2 = |\mu|^2 + m_{H_d}^2
\label{scalarmass}
 \eq
and $m_{H_u}^2$ and $m_{H_d}^2$ are the Higgs soft masses. Without
loss of generality we can take the soft parameter $m_{ud}^2$ to be
real. It follows that the tree-level MSSM Higgs potential is CP
conserving (even though the full MSSM Lagrangian violates CP).
The massive Higgs particles are eigenstates of this approximate
CP.  The light Higgs $h$ and the heavier Higgs $H$ are CP even,
and the Higgs $A$ is CP odd.  In addition there is a massive
charged Higgs $H^\pm$.

The tree level potential depends on the known $SU(2)_L \times
U(1)_Y$ gauge couplings $g$ and $g^{\prime}$ and three unknown
real mass squared parameters, $\widetilde m_{H_u}^2$, $\widetilde
m_{H_d}^2$, and $m_{ud}^2$. It is convenient to parameterize the
observables not in terms of these three real parameters, but in
terms of three other quantities.  Two of them are
 $$ v_u = |\langle H_ u\rangle| = v \sin \beta $$
\beq
v_d=  |\langle H_d \rangle| = v \cos \beta
\eeq
and the third is the physical mass of the CP odd Higgs, $m_A$.
Since the expectation value $v$ is known, this leaves
$m_A$ and $\tan \beta$ as the two unknown parameters
describing the MSSM Higgs sector.
Also, instead of using the gauge couplings $g$ and $g^\prime$, we
will write expressions
in terms of the gauge boson masses,
$m_Z^2 = {1 \over 2} (g^2 + g^{\prime 2}) v^2$,
$m_W^2 = {1 \over 2} g^2 v^2$.

Then, a straightforward computation leads to the masses
 $$
m_{h,H}^2 = {1 \over 2} \left[
  m_Z^2 + m_A^2 \mp \sqrt{
  (m_A^2 - m_Z^2)^2 + 4 m_A^2 m_Z^2 \sin^2 2 \beta } \right]
 $$
 \beq m_{H^{\pm}}^2 = m_A^2 +
 m_W^2.~~~~~~~~~~~~~~~~~~~~~~~~~~~~~~~~ ~~~~~~~~~~~~~~~~~~
 \label{higgstreeg}
 \eeq
We note that for $g=0$ the expression (\ref{higgstreeg}) becomes
$m_{H^\pm}=m_A$. In this case the Higgs potential has a custodial
$SU(2)_C$ symmetry under which $H^\pm$ and $A$ form a triplet, as
discussed in the Apppendix \cite{gerard}.  While of little
importance in the MSSM, the further breaking of this symmetry in the
presence of new physics is distinctly more interesting.

It is worth noticing that the last expression in
(\ref{higgstreeg}), as well as
 \beq
 m_h^2 + m_H^2 =  m_Z^2 + m_A^2
 \label{tracerelation}\eeq
are independent of $\tan \beta$ and therefore provide interesting
one parameter
tests of the MSSM (with of course proper quantum corrections
taken into account).

It follows from (\ref{higgstreeg}) that $m_h \le m_Z$, which is
incompatible with the LEPII bound.   Although radiative
corrections and various operator corrections to the Higgs
potential discussed below can resolve this contradiction, this
problem is less severe when this tree level inequality is nearly
saturated.  This is the case for large $\tan \beta$. Therefore, we
will expand (\ref{higgstreeg}) in powers of $ \eta \equiv \cot
\beta .$

In treating $\eta$ as small, we should decide what to hold fixed.
In the standard decoupling limit, $m_A \to \infty$.  We prefer,
instead, to take large $\tan \beta$ with fixed $m_A$.  This limit
is more physical because the spectrum remains finite; indeed, if
we wish to avoid large, highly tuned hierarchies, we don't expect
$m_A$ to be vastly different than the $W$ and $Z$ boson masses.
Also, in this limit the parameters in (\ref{VMSSMtree}) remain
finite while $m_{ud}^2$ is taken to zero. In this limit the
potential (\ref{VMSSMtree}) acquires a $U(1)_{PQ}$ symmetry, so
the limit $\eta \to 0$ becomes technically natural. A
straightforward calculation leads to
 $$ m_h^2 \simeq m_Z^2 - { 4m_Z^2 m_A^2  \over
   m_A^2 - m_Z^2 }\eta^2 + {\cal O}(\eta^4)
$$
$$ m_H^2 \simeq m_A^2 + { 4m_Z^2 m_A^2 \over
   m_A^2 - m_Z^2 }\eta^2+ {\cal O}(\eta^4)
 $$
 \beq m_{H^{\pm}}^2 = m_A^2 + m_W^2 .~~~~~~~~~~~~~~~~~~~~
 \eeq
Note that $H$ and $A$ become degenerate in the large $\tan \beta$
limit.  They form a complex boson which carries $U(1)_{PQ}$
charge.  Also, the expression for $m_{H^\pm}$, which is
independent of $\eta$, is not corrected.

% ----------------------------

\subsection{Radiative corrections}

Radiative corrections are known to significantly change the
spectrum \cite{oneloophiggs,cponeloop}.  They are most important for the
light Higgs $h$, for two reasons.  First, the tree level mass is
small (because it is proportional to the small gauge couplings).
Second, the loop corrections are proportional to four powers of
the top Yukawa coupling. These radiative corrections can avoid a
contradiction with the LEPII bound.  The most important effect is
that of virtual loops of tops and stops. The leading one-loop top
and stop corrections to $m_h^2$ in either the large $\tan \beta$
or Higgs decoupling limits including stop mixing effects with
arbitrary phases are \cite{oneloophiggsphases,cpviolating}
$$
\delta_{1-{\rm loop}}\ m_{h}^2 \simeq  {12 \over 16 \pi^2} {m_t^4
\over v^2} \left[ \ln \left( {m_{\widetilde{t}_1}
m_{\widetilde{t}_2} \over m_t^2} \right) + { |X_t|^2  \over
m_{\widetilde{t}_1}^2 - m_{\widetilde{t}_2}^2 }
   \ln \left( {m_{\widetilde{t}_1}^2 \over  m_{\widetilde{t}_2}^2}
   \right)   \right. ~~~~~~~~
$$
\beq
 \left.
  ~~~~~~~~~~~~~~~~~~
 +  {1 \over 2} \left(
{ |X_t|^2 \over  m_{\widetilde{t}_1}^2 - m_{\widetilde{t}_2}^2  }
    \right)^2
  \left( 2 - { m_{\widetilde{t}_1}^2 + m_{\widetilde{t}_2}^2 \over
               m_{\widetilde{t}_1}^2 - m_{\widetilde{t}_2}^2 }
                \ln \left( {m_{\widetilde{t}_1}^2 \over  m_{\widetilde{t}_2}^2} \right)
               \right)
          \right]
  \label{mhoneloop}
 \eq
where $v = (2^{3/2} G_F)^{-1/2} \simeq 174$ GeV, $X_t = A_t -
\mu^* \eta $ is the stop mixing parameter, and
$m_{\widetilde{t}_{1,2}}$ are the stop physical masses (in the
limit of large $\tan \beta=1/\eta$ with $m_A$ held fixed, $X_t$ may
be replaced by $A_t$).

The current bound from LEPII on the mass of a Standard Model like
Higgs boson of $m_{h_{\rm SM}} \gsim 114$ GeV \cite{LEPhiggsbound}
can be accommodated within the MSSM in certain regions of
parameter space. First, the tree-level contribution is maximized
at moderate to large $\tan \beta$. For $\tan \beta \gsim 10$ the
tree-level contribution is already within 2 GeV of its maximum
value of $m_Z$.   Moreover, the top and stop loop correction can
be substantial. If stop mixing is small, $|X_t/
m_{\widetilde{t}_{1,2}}|^2 \ll 1$, the correction depends only on
the logarithm of the stop masses, and the stops must be rather
heavy in this small mixing limit. Including the full set of
two-loop corrections with typical parameters for the remaining
MSSM parameters, the LEPII Higgs mass bound requires that
$m_{\widetilde{t}_{1,2}} \gsim 1000$ GeV in the small stop mixing
limit \cite{feynhiggs,rouven}. However, from (\ref{mhoneloop}), one sees
that if stop mixing is large, the stop loop correction can be
sizeable, allowing much lighter stops \cite{feynhiggs,choi,stopmixing}. At
one-loop the stop mixing contribution is maximized for $|X_t/
m_{\widetilde{t}}|^2 \simeq 6$ where $m_{\widetilde{t}} = {1 \over
2}( m_{\widetilde{t}_1} + m_{\widetilde{t}_2})$. Including the
full set of two-loop corrections with the maximum correction from
stop mixing, the LEPII Higgs mass bound is consistent with
$m_{\widetilde{t}_1} \gsim 100$ GeV \cite{feynhiggs,rouven},
as required by
the LEPII direct stop search.  While such large mixing is not
inconceivable, it is typically hard to obtain in specific
mediation schemes, and it generally arises at low energies under
renormalization group evolution only from rather special points in
parameter space \cite{rouven}.

\section{The Higgs Sector Window to New Physics}
\label{sec:window}

% --------

\subsection{Operator Analysis}
\label{sec:operatoranalysis}

The effects of new physics at a sufficiently large mass scale,
$M$, can be encapsulated in new operators. The magnitude of the
interactions arising from these operators, in turn, should be
organized in inverse powers of the heavy mass scale $M$.  As we
explained in the introduction, the power of $1/M$ is related to
the {\it effective dimension} of the operator.

In the supersymmetric limit the leading interactions arise from a
single operator presented below, which is suppressed by a single
power of $1/M$. In terms of component fields this operator
generates both renormalizable as well as non-renormalizable
interactions.  The leading component operator is of dimension four
but its coefficient is suppressed by the dimensionless combination
 \beq
  \epsilon \sim {|\mu| \over M}
 \eeq
where $\mu$ is the coefficient of the term $H_uH_d$ in the
superpotential. Therefore, this operator has effective dimension
five.

For operators which include supersymmetry breaking, an important
consideration is the scale of supersymmetry breaking, $m_{\rm
SUSY}$, within the new physics sector. We will assume
 \beq
 m_{\rm SUSY} \sim |\mu|
 \eeq
in this section. Since $m_{\rm SUSY} \ll M$, the effects of
supersymmetry breaking may be described through supersymmetric
operators which contain spurions with supersymmetry breaking
auxiliary component expectation values. The leading interactions
which include supersymmetry breaking arise from a single operator
presented below, suppressed by a one power of
 \beq
 \epsilon \sim {m_{\rm SUSY}\over  M}.
 \eeq
In terms of component fields this operator generates a
renormalizable interaction, but is suppressed by one power of
$1/M$, so is of effective dimension five.  It is important to
stress that the explicit supersymmetry breaking due to these
operators is not soft. These operators are of dimension four and
the breaking is hard in the low energy effective theory. Yet, this
is perfectly consistent because their coefficients are suppressed
by powers of $M$.

Let us be more explicit, starting with the supersymmetric
operators. Consider first superpotential interactions.  The most
general superpotential for the MSSM Higgs sector up to dimension
five is
 \beq \int d^2\theta \left( \mu H_u H_d
 + { \lambda \over M} (H_u H_d)^2 \right).
 \label{genHsuper}
 \eq
After eliminating Higgs auxiliary fields with the leading order
canonical Kahler potential, to leading order in $\mu / M$ this
gives a correction to the renormalizable MSSM Higgs potential
 \beq
\delta_1 V = 2 \epsilon_1 ~
   H_u H_d ~( H_u^\dagger H_u + H_d^\dagger H_d )  ~+~{\rm h.c.}
 \label{vepsilon1} \eq
 where
  \beq \epsilon_1 \equiv {\mu^* \lambda \over M}. \eeq

The superpotential (\ref{genHsuper}) also gives a dimension five
correction to the Higgs-Higgsino Lagrangian interaction:
 \beq
   { \epsilon_1 \over \mu^*} \left[ 2 (H_u H_d)
  (\widetilde{H}_u \widetilde{H}_d)
  +2 (\widetilde{H}_u {H}_d   )
  (H_u \widetilde{H}_d   )
 + (H_u \widetilde{H}_d ) (H_u \widetilde{H}_d )
 + (\widetilde{H}_u H_d) (\widetilde{H}_u H_d )  \right]  ~+~{\rm h.c.}
\label{higgshiggsino}
 \eq
where here round parenthesis indicates $SU(2)_L$ singlet.
Substituting the scalar
Higgs expectation values in this expression, the neutralino
and chargino masses are corrected at order $\epsilon_1$.

We should also consider Kahler potential interactions. Gauge
invariance requires that all such operators which involve only
Higgs fields are functions of an even number of fields. The lowest
dimension non-trivial Kahler potential operators beyond the
leading kinetic terms are therefore dimension six with four Higgs
fields and are suppressed by $1/M^2$. There are several such
operators and their effects are typically smaller than those of
(\ref{vepsilon1}) and (\ref{higgshiggsino}) which are suppressed
by one power of $1/M$.  We will discuss some of them below.

The second class of operators involve supersymmetry breaking.
These effects may be included by considering superpotential and
Kahler potential operators with additional powers of
spurion superfields with auxiliary expectation values.
For definiteness we consider here the phenomenologically
interesting case of an $F$-term auxiliary expectation value.
This may be represented by a dimensionless
chiral superfield spurion
 \beq
 {\cal Z} = \theta^2 m_{\rm SUSY},
 \eeq
where $m_{\rm SUSY}$ is the supersymmetry breaking scale.
Superpotential operators give non-vanishing interactions only with
a single power of the spurion ${\cal Z}$. The dimension five
operator similar to that of (\ref{genHsuper}), but with a
spurion field
 \beq \int d^2 \theta {\cal
 Z} ~ { \lambda \over M} (H_u H_d)^2 \label{WZHHHH} \eq
gives a holomorphic correction to the renormalizable MSSM Higgs
potential:
 \beq
 \delta_2 V =  \epsilon_2 ~(H_u H_d)^2 ~+~{\rm h.c.}
 \label{genAcorrection} \eq
where
 \beq \epsilon_2 = - {m_{\rm SUSY} \lambda \over M}.
  \eeq
This correction is also of dimension four, but effective dimension
five.

Kahler potential operators give non-vanishing interactions with
either one or two powers of the spurion ${\cal Z}$. To zeroth
order in $1/M$ (as in the MSSM) only the Kahler potential kinetic
terms can contribute with one or two powers of the spurion field.
As is well known, the terms with one spurion like\footnote{Here
and in the rest of this note $e^V$ represents the exponential of
the gauged superfield in the appropriate representation.}
 \beq
 \int d^4\theta ({\cal Z} + {\cal Z}^{\dagger})~
   X^{\dagger} e^V X
   \label{KZ}
 \eq
where $X$ is any of the MSSM fields, can be removed by making
holomorphic field redefinitions
 \beq
 X \rightarrow \left( 1 - {\cal Z} \right)X.
 \eq
This restores the canonical kinetic terms for $X$ and generates
various operators which are already included in the MSSM
Lagrangian and the operator (\ref{WZHHHH}). Therefore, the
operator (\ref{KZ}) is not an independent operator. Terms which
are second order in the spurion
 \beq
 \int d^4\theta {\cal Z Z}^\dagger X X^\dagger
 \label{KZZ}
 \eq
lead to standard soft masses.

Proceeding to first order in $1/M$, i.e.\ effective dimension
five, we easily see that there are no terms which involve only the
Higgs fields.  Allowing for couplings to other fields,
there are operators such as\footnote{We
thank L.~Randall for a useful discussion about this point.}
 \beq
  \int d^4\theta {1 \over M} \left( 1+ {\cal Z} + {\cal
  Z}^{\dagger}  + {\cal Z Z}^{\dagger}\right)~ \left( H_d^\dagger
  Q u^c + H_u^\dagger Q d^c   + H_u^\dagger L e^c
  \right).
   \label{dimfK}
 \eq
Here we have suppressed order one coefficients, so even with a
single generation, there are 12 independent operators; allowing
for generation indices on the quark and lepton doublets $Q$ and
$L$, the anti-up quarks $u^c$, the anti-down quarks $d^c$, and the
anti-leptons $e^c$,  there are many independent operators. Such
operators are potentially dangerous because they might lead to
FCNC.  We will assume that the short distance physics is such that
if these operators are generated, their coefficients are
sufficiently small.  It should be stressed, however, that this
assumption is not that strong, because a similar assumption is
standard already in the zeroth order terms (\ref{KZZ}).

In addition there are various dimension five superpotential
operators which violate baryon
and lepton number \cite{invariants}.
We do expect such operators to be present and
to lead to neutrino masses and proton decay.  But we expect their
scale to be much larger than the scale $M$ we contemplate here,
which we assume to be in the TeV range.

The Higgs interactions (\ref{vepsilon1}) and
(\ref{genAcorrection}) and Higgs-Higgsino interactions
(\ref{higgshiggsino}) exhaust all the MSSM Higgs sector effective
dimension five operators that arise either in the supersymmetric
limit or from chiral superfield spurions.
None of these component interactions
arise at tree-level for the renormalizable MSSM Higgs sector with
general soft supersymmetry breaking. All these interactions
violate the continuous $U(1)_{PQ}$ symmetry under which both the
$H_u$ and $H_d$ superfields have unit charge. In the MSSM this
symmetry is, however, already broken by the superpotential Higgs
mass $\mu$-term. So the requisite breaking of $U(1)_{PQ}$ by the
new physics sector need not necessarily lead to any additional
suppression of the effective dimension five interactions. This can
be true even in the case that $m_{ud}^2$ in the tree-level
potential (\ref{VMSSMtree}) is small since this term breaks
different discrete symmetries than the $\mu$-term or effective
dimension five operators.

There are a large number of interactions with MSSM particle
content at effective dimension six. A subset of these have
the property that their component expansion involves terms which
depend only on $H_u$. These play an important role in the
discussion in the next subsection of the leading effects which can
modify the light Higgs boson mass for sufficiently small $\eta$,
since their effects are independent of $\eta$ in this limit.
Effective dimension six component interactions which involve only
the scalars in $H_u$ come entirely from the dimension six Kahler
potential operators
$$
 \int d^4\theta  {1 \over M^2} \left[
 \xi_1 (H_u H_d)^{\dagger} (H_u H_d)  +
 \xi_2 (H_u e^V H_u^{\dagger}) (H_d e^V H_d^{\dagger})+
 \xi_3 (H_u^{\dagger} e^V H_u )^2
 \right.
$$
\beq
 \left. ~~~~~
 + \left( \xi_4 {\cal Z} (H_u H_d)^{\dagger}  (H_u^{\dagger} e^V H_u )
 + {\rm h.c.} \right)
 + \xi_5  {\cal Z}^{\dagger} {\cal Z} (H_u^{\dagger} e^V H_u )^2
 \right].
 \label{kahlersix}
 \eq
Using $F_{H_d} \simeq -(\mu H_u)^\dagger$ these operators modify
the scalar potential for $H_u$ by operators of dimension four and
six and also modify its kinetic term.

It is important to stress again that the effective operator
analysis given above for the MSSM Higgs sector differs
significantly from that of a non-supersymmetric theory with the
same field content. Here the analysis is best organized in terms
of the effective dimension of component interactions. In contrast,
the analysis for a generic non-supersymmetric theory is most
naturally organized in terms of the operator dimension of the
interactions.

The effective operator analysis presented above, including the
effects of supersymmetry breaking by $F$-term auxiliary
expectation values through the spurion ${\cal Z}$, may be extended
to include $D$-terms. In most microscopic models of supersymmetry
breaking such auxiliary fields are suppressed and unimportant, and
therefore they are somewhat less familiar.  However, for
completeness it is interesting to classify the effects of such
fields in the effective operator analysis.

A $D$-term expectation value resides in a gauge vector superfield.
If MSSM fields carry the corresponding charge, the relevant
spurion is a standard vector field of dimension zero and its
contribution to the scalar masses generally requires a suppression
of the $D$-term vevs.  If instead all the MSSM fields are
uncharged under the associated gauge symmetry, we may use a
spurion vector superfield of dimension $-1$, and the effects of
$D$-term breaking are represented by a spinor chiral superfield
strength spurion of scaling dimension ${1 / 2}$
 \beq
 {\cal W}_{\alpha} = \xi_{\rm SUSY} \theta_{\alpha} \qquad {\rm
 with} \qquad \xi_{\rm SUSY}\sim m_{\rm SUSY}.
 \label{Wspurion} \eq
The leading coupling of this spurion to the MSSM is the dimension
three operator
 \beq M \int d^2 \theta~ {\cal W}^{\alpha}
 W_{\alpha}^{(1)} = M \xi_{\rm SUSY} D^{(1)}  \eq
where $W_{\alpha}^{(1)}$ is the $U(1)_Y$ hypercharge spinor
superfield strength and $D^{(1)}$ the hypercharge $D$-term.
This is the standard supersymmetric Fayet-Iliopoulos
term for $U(1)_Y$ hypercharge.   Since for $\xi_{\rm
SUSY}\sim m_{\rm SUSY}$ this term is too large, its coefficient
must be suppressed.  This is easy to arrange in specific
microscopic models.

Up to redundancies due to the lowest order equations of motion
and holomorphic field redefinitions
there are no dimension four couplings of the spurion
(\ref{Wspurion}) to MSSM fields. At dimension five there are two
couplings which involve Higgs sector fields (again, we limit
ourselves to baryon and lepton number conserving operators)
 $$
 {1 \over M} \int d^2 \theta ~ {\cal W}^{\alpha}  H_u H_d
 W_{\alpha}^{(1)}= {\xi_{\rm SUSY} \over M} \left[
H_u H_d D^{(1)} + {i \over \sqrt{2} } \left(
  H_u \psi_{H_d} + \psi_{H_u} H_d \right)   \lambda^{(1)}
 \right] $$
 \beq
 {1 \over M} \int d^2 \theta ~ {\cal W}^{\alpha}  H_u  H_d
 W_{\alpha}^{(2)}= {\xi_{\rm SUSY} \over M} \left[
 H_u  H_d D^{(2)} + {i \over \sqrt{2} } \left(
  H_u \psi_{H_d} + \psi_{H_u} H_d \right)   \lambda^{(2)}
 \right]
 \label{WHHW}\eq
where $W_{\alpha}^{(2)}$ is the $SU(2)_L$ spinor superfield
strength.  Using the lowest order values of the $D$-terms, the
bosonic terms give corrections to the Higgs potential of the form
$H_u H_d (H^{\dagger}_u H_u)$ and $H_u H_d (H^{\dagger}_d H_d)$.
These are similar to the effective dimension five terms
(\ref{vepsilon1}). However, unlike those, in (\ref{WHHW}) the
relative coefficients of the two component interactions are not
related. In addition, the Higgs potential correction
(\ref{vepsilon1}) is related to the Higgs-Higgsino interaction
(\ref{higgshiggsino}), whereas in (\ref{WHHW}) there are no such
component interactions.  Instead, there are new dimension four but
effective dimension five Higgs-Higgsino-gaugino interactions.
Substituting the Higgs expectation values, these interactions
correct the chargino and neutralino masses at order $1/M^2$.

% ---------------------------------------------------------------------------------------

\subsection{The Higgs Spectrum}
\label{sec:lighthiggs}

We now consider the effect of the operators above on the Higgs
masses.  We add to the MSSM potential (\ref{VMSSMtree}) the two
terms $\delta_1V$ (\ref{vepsilon1}) and $\delta_2V$
(\ref{genAcorrection}).  Again, we express the coefficients of the
quadratic terms in terms of $v$, $\tan \beta$ and $m_A$ and expand
the answers to leading order in $\epsilon$:
 $$
 \delta_\epsilon m_h^2 = 2v^2\left(\epsilon_{2r} + 2 \epsilon_{1r}
 \sin(2\beta) + {2 \epsilon_{1r}(m_A^2 + m_Z^2) \sin(2\beta) -
 \epsilon_{2r} (m_A^2-m_Z^2) \cos^2(2\beta) \over
 \sqrt{(m_A^2-m_Z^2)^2 + 4m_A^2 m_Z^2 \sin^2(2\beta)}} \right)$$
  \beq
   \simeq
 {16 m_A^2  \over m_A^2 - m_Z^2} v^2\eta \epsilon_{1r}
 +\CO(\eta^2 \epsilon)
 ~~~~~~~~~~~~~~~~~~~~~~~~~~~~~~~~~~~~~~~~~~~~~~~~~~~~~~~~~
 \label{masseighre}
 \eeq
$$
 \delta_\epsilon m_{H}^2 =2v^2\left(\epsilon_{2r} + 2
 \epsilon_{1r}
 \sin(2\beta) - {2 \epsilon_{1r}(m_A^2 + m_Z^2) \sin(2\beta) -
 \epsilon_{2r} (m_A^2-m_Z^2) \cos^2(2\beta) \over
 \sqrt{(m_A^2-m_Z^2)^2
 + 4m_A^2 m_Z^2 \sin^2(2\beta)}} \right)
 $$
 \beq
 \simeq  4 v^2 \epsilon_{2r}
 - {16 m_Z^2 \over m_A^2 -m_Z^2}  v^2 \eta \epsilon_{1r}
 +\CO(\eta^2 \epsilon)
 ~~~~~~~~~~~~~~~~~~~~~~~~~~~~~~~~~~~~~~~~~~~~
  \label{deltamHe}\eeq
 \beq
 \delta_\epsilon m_{H^\pm}^2 = 2 \epsilon_{2r} v^2
 ~~~~~~~~~~~~~~~~~~~~~~~~~~~~~~~~~~~~~~~~~~~~~~~~~~~~~~~~~~~~~~~~~~~~
 ~~~~~~~~~~~~~~~
 \label{deltamHc}
 \eeq
where $\epsilon_{1r}$ and
$\epsilon_{2r}$ are the real parts of $\epsilon_1$ and
$\epsilon_2$.

We would like to make some simple comments
about these expressions:
\begin{enumerate}
 \item These expressions, at first order in $\epsilon$, depend
only on the real parts of $\epsilon_{1,2}$.   This can be
understood as follows.  At the zeroth order in $\epsilon$ the
problem is CP invariant.  At first order in $\epsilon$, the effect
of ${\rm Im}\, \epsilon_{1,2}$ is to mix CP even and odd states.
But the spectrum is only affected in second order. Interactions
(decays, dipole moments, and the like) will be affected at order
$\epsilon$.
 \item The expressions (\ref{masseighre}) and (\ref{deltamHe})
receive also corrections of higher order in $\epsilon_{1,2}$ (from
the potential evaluated to first order in $\epsilon)$.  For
example,
 \beq
 \delta_{\epsilon^2}m_h^2=-{16 \epsilon_{1r}^2+48 \epsilon_{1i}^2
 - 4 {m_Z^2 \over m_A^2} \epsilon_{2i}^2
  \over m_A^2 -m_Z^2}~
  v^4+ {\cal O}(\eta \epsilon^2),
 \label{ordereps}
 \eq
where we see that, as we commented above,
the order $\epsilon^2$
corrections depend on ${\rm Im}\ \epsilon_1=\epsilon_{1i}$ and
${\rm Im}\ \epsilon_2=\epsilon_{2i}$.  We should point out that
with CP violation the limits $\epsilon\to 0$ and $\eta \to 0$ are
singular and more care is needed in the expansions.  Note that the
operator (\ref{genHsuper}) also leads to a sixth order term in the
potential which is suppressed by $\epsilon_1^2$.  This term does
not contribute to (\ref{ordereps}) because it includes at least
two factors of $H_d$.
 \item Although the expressions for
both (\ref{masseighre}) and (\ref{deltamHe}) receive higher order
corrections in $\epsilon_{1,2}$ (as can be seen in
(\ref{ordereps})), for $\epsilon_{1i}=\epsilon_{2i}=0$ the sum
$m_h^2 + m_H^2$ (\ref{tracerelation}) and $m_{H^\pm}$
(\ref{deltamHc}) are given exactly by the order $\epsilon $
correction above. This comment, however, is of only academic
interest, because corrections to the potential of higher effective
dimension will modify both $m_h^2 + m_H^2$ and $m_{H^\pm}$.
 \item The expression for $\delta_\epsilon m_{H^\pm}^2$
(\ref{deltamHc}) is very simple and is independent of $\tan \beta
$ and $\epsilon_1$. This can be understood as a consequence of an
approximate global $SU(2)_C$ custodial symmetry which is discussed
in the Appendix.
 \item Focusing on the leading order in $\epsilon$ and assuming
 that $\eta$ is very small, the corrections to the heavy Higgs
 masses $m_H$ and $m_{H^\pm}$ are nonzero and
depend only on $\epsilon_{2r}$.  Since the corrections to these
two masses depend on only one real number, experimental values of
these masses will over determine it (or bound it) and will serve
as a nontrivial test of the existence of this operator. In
particular, the $\epsilon_2$ independent relation
 \beq
 2m_{H^{\pm}}^2 = m_H^2 + m_A^2 + 2 m_W^2 + {\cal O}( \eta^2,\eta
 \epsilon, \epsilon^2 )
 \eq
serves as a test of the BMSSM effective dimension five
parameterization (with of course proper quantum corrections taken
into account).
 \item The corrections to the light Higgs mass $m_h^2$ are
of order $ \epsilon \eta$ and hence they are negligible for very
small $\eta$.  If $\eta$ is such that this contribution is
measurable, it will determine the value of $\epsilon_{1r}$.
\end{enumerate}

One of the motivations for our analysis is to classify effects
which may lift the light Higgs mass.  We now see that for $\eta$
parameterically very small the effect of terms at first order in
$\epsilon$ is negligible. More precisely, if $\eta \gg \epsilon$
then the result (\ref{masseighre}) is useful. However, if $\eta
\lsim \epsilon$, then the order $\epsilon^2 $ corrections are
comparable, and perhaps even larger than the order $\epsilon \eta$
and they are parametrically the leading correction to $m_h^2$.

First, let us ignore the various order $\epsilon^2$ corrections.
Then, the total shift of the light Higgs mass depends on both the
radiative corrections (\ref{mhoneloop}) as well as the corrections
(\ref{masseighre}) from the operators discussed above. As a
conservative numerical example of the magnitude of the radiative
corrections, stop squark masses of $m_{\tilde{t}_{1,2}} \simeq
300$ GeV with small mixing, $X_t \simeq 0$, yield a Higgs mass at
moderate to large $\tan \beta$ of $m_h \simeq 100$ GeV. Even in
this small mixing case, and with $m_A \gg m_Z$, the additional
correction (\ref{masseighre}) can accommodate the LEPII Higgs mass
bound of $m_h \gsim 114$ GeV for $\eta \epsilon_{1r} \gsim 6
\times 10^{-3}$. This could be achieved for example with $\eta
\sim 0.1$ and $\epsilon_{1r} \sim 0.06$ which for $\mu \sim 300$
GeV would correspond to a scale in the operator (\ref{genHsuper})
of $M/ \lambda \sim 5$ TeV. The leading correction
(\ref{masseighre}) grows with decreasing $m_A$, and so is slightly
larger away from the Higgs decoupling limit.

Next, consider the limit $\eta \ll 1$, where $\epsilon^2$
corrections to the light Higgs mass are important.  In this case,
$h$ arises predominantly from $H_u$.  Therefore, we should examine
the corrections to the potential and the kinetic term of $H_u$.
The relevant operators are enumerated in (\ref{kahlersix}).  At
order $\epsilon^2$ we need to consider the leading order effect of
these operators as well as the order $\epsilon^2$ corrections to
the masses computed with the order $\epsilon$ correction to the
potential as in (\ref{ordereps}).  Since we are interested only in
the zeroth order in $\eta$ we can absorb all the unknowns in one
number $\epsilon_3 \sim \epsilon$ and write (for real
$\epsilon_{1,2}$)
 \beq
 \delta_{\epsilon^2} m_h^2 = -{16 v^4 \over m_A^2 -m_Z^2}~
 \epsilon_{1r}^2+ \epsilon_3^2 v^2.
 \label{deltaepss}
 \eeq
For the small mixing example given above, the LEPII Higgs mass
bound can be accommodated with the second order corrections
(\ref{deltaepss}) for $\epsilon_3 \gsim 0.3$. With $m_{\rm SUSY}
\sim 300$ GeV and $\epsilon_3 \sim m_{\rm SUSY} / M$ this
corresponds to $M \sim 1$ TeV.

In addition to modifications of the scalar Higgs masses, the
operator (\ref{genHsuper}) also modifies the Higgsino masses
through the interactions (\ref{higgshiggsino}).
The first and second terms in
(\ref{higgshiggsino}) with scalar Higgs expectation values
correct the charged and neutral Higgsino Dirac masses.
The third and fourth terms in
(\ref{higgshiggsino}) with scalar Higgs expectation values give
rise to neutral Higgsino Majorana masses which are absent
in the tree-level neutralino mass matrix.
Precision fits to both masses and couplings of neutralinos and
charginos would be sensitive to the dimension five Higgs-Higgsino
interactions.
It is important to note that the interactions
(\ref{higgshiggsino}) are all proportional to a single coupling,
$\epsilon_1$, which is the same as the coupling affecting the
Higgs mass.
They are not the most general component couplings at
this operator dimension which would arise among the Bino, Wino,
Higgsinos, and Higgs scalars in a general non-supersymmetric
theory.
So a precision fit to the BMSSM with the two effective
dimension five operators is still highly overconstrained.

% ------------------------------------------------------------

% ------------------------------------------------------------
\section{Microscopic Models}
\label{sec:microscopicmodels}

Various types of dynamics might give rise to the various operators
we have enumerated above.  The most widely explored are the NMSSM
\cite{nmssm}, the addition of a singlet to the MSSM, and the
possibility of additional gauge interactions at low scales
\cite{gaugeref,gaugerefb}. Here we analyze both types of models in
the framework we have described above.  Although we will not
discuss it here, it is clear that the dynamics leading to these
operators can involve new strongly coupled sectors in the TeV
range.

\subsection{Adding a Singlet:  The NMSSM}

The most studied extension of the MSSM is a model with an
additional singlet, $S$, the NMSSM \cite{nmssm}. This model is
usually motivated as an explanation of the $\mu$ term, or, more
generally, to avoid the appearance of any parameter with
dimensions of mass in the low energy Lagrangian.  As a result, one
usually writes:
 \beq \int d^2\theta \left(\lambda_S S H_u H_d
 + {\lambda' \over 3} S^3 \right).
 \label{nmssmw} \eeq
This structure can be enforced, for example, by a discrete $R$
symmetry. Supersymmetry breaking (e.g.\ through the judicious
choice of soft mass terms for $S$ and $H$) then leads to an
expectation value for $S$, which in turn generates an effective
$\mu$ term.

In a model of this sort, the $S$ field is unlikely to be
significantly more massive than the Higgs field, and it does not
make sense to integrate it out to obtain a local action.  The
problem is that the couplings $\lambda_S$ and $\lambda'$ cannot be
too large, at least if the model is to remain perturbative up to
high energies; typically one requires $\lambda_S < 0.7$.  At the
same time, one can't make $\lambda_S$ very small, if one is to
obtain a substantial Higgs mass.  Given that $\langle S \rangle =
\mu/\lambda_S$, any {\it supersymmetric} mass term for $S$ is
given by $M_S = 2 \lambda' \mu/\lambda_S$, and this can't be much
larger than $\mu$.

It has been appreciated for a long time that dimensionful
parameters such as $\mu$ and a supersymmetric mass term for $S$
can arise through couplings to some non-perturbative dynamics.
More specifically, following
\cite{yanagida,lutysundrum,dfs,aharonyseiberg,dinemason} if one has some
new, pure gauge theory, with a characteristic scale $\Lambda$,
then couplings such as
 \beq
 {W_\alpha^2 \over M_0^2}(a_1 H_u H_d + {a_2 \over 2}S^2) + \lambda_S
 S H_u H_d. \label{dynamicalmu} \eeq
where $a_1$ and $a_2$ are numerical constants and $M_0$ is some
heavy scale, give rise to $\mu$ and $M_S$ of order
$\Lambda^3/M_0^2$. As in the case of the conventional NMSSM
(\ref{nmssmw}), this structure can be enforced by discrete
symmetries (in which case, necessarily, a ``bare" mass for Higgs
or $S$ are forbidden).

We work, then, with a model with superpotential:
 \beq \int d^2\theta \left( \mu H_u
 H_d + {1 \over 2} M_S S^2 + \lambda_S S H_u H_d \right
 )  .\eq
If $M_S \gg \mu$ (which arises in (\ref{dynamicalmu}) if $a_2 \gg
a_1$), the heavy singlet may be integrated out through its
holomorphic equation of motion:
 \beq S = - {\lambda_S \over M_S} H_u H_d.
 \label{singleteom} \eq
Here we neglected corrections involving covariant derivatives
because they lead to higher order corrections to the effective
action.  In the supersymmetric limit this gives a tree-level
Kahler potential and superpotential
 \beq \int d^4\theta \left( H_u^{\dagger} e^V
H_u + H_d^{\dagger} e^V H_d + \left| { \lambda_S \over M_S}
\right|^2
 (H_u H_d)^{\dagger} (H_u H_d) \right)
 \label{singletKeff}
 \eq
  \beq \int d^2 \theta \left(
  \mu H_u H_d - {\lambda_S^2 \over 2 M_S} (H_u H_d)^2 \right).
\label{singletWeff} \eq

{}From (\ref{singletWeff}) we can read off the coefficient of the
operator of (\ref{genHsuper}):
 \beq {\lambda \over M} = -{\lambda_S^2 \over 2 M_S}\qquad ;
 \qquad \epsilon_1 = -{\mu^* \lambda_S^2 \over 2
 M_S}.
 \label{eps1singlet}
 \eeq
Supersymmetry-breaking can be described to leading order by a
spurion coupling
 \beq \int d^2 \theta~ {\cal Z}{1 \over 2}M_S S^2 \eeq
After integrating out the singlet with this soft breaking, the
operator (\ref{WZHHHH}) is generated with %(\ref{genAcorrection})
 \beq
 \epsilon_2 = {m_{\rm SUSY} \lambda_S^2 \over M_S}.
 \eeq

The first order corrections in $\epsilon$ can be substantial. For
example, taking $\tan \beta = 4$, the stop soft masses 300 GeV,
and some sample values of $X_t$, the mass of the lightest Higgs in
the MSSM (i.e.\ the mass including radiative corrections but
excluding the contributions of the higher dimension operators),
$m_{h\ \rm MSSM}$, is given in the second column of the table
below.  Adding the singlet $S$, with $M=5\mu$, $\lambda_s = 0.7$
and different values of $\epsilon_2$, the mass of the light Higgs,
$m_h$, is readily pushed above the LEPII bound:

\vskip .8cm

\begin{center}
\begin{tabular}{|c|c|c|c|}
  \hline
%  $\tan \beta$ & $X_t/m_{\tilde t}$ & $m_{h\ \rm MSSM}$ & $|\mu/M| $
%  & $\epsilon_2$ & $m_h$ \\
  $X_t/m_{\tilde t}$ & $m_{h\ \rm MSSM}$ (GeV)
  & $\epsilon_2$ & $m_h$ (GeV)\\
  \hline
  %3 & 0 & 84 & 1/5 & 0 & 109 \\
  %3 & 1 & 91 & 1/5 & 0 & 115\\
 0 & 91 &  $0$ & 120 \\
 0 & 91 &  $\epsilon_1$ & 124  \\
 1 & 98 &  $0$ & 126 \\
 1 & 98 &  $\epsilon_1$ & 129 \\
  \hline
\end{tabular}
\end{center}

\vskip .5cm

In this model, we  can assess the validity of the expansion in
$\epsilon_i$ and $\eta$. There are a number of terms at second
order in $\epsilon$ which can increase or decrease the mass of
$h$.  One set of contributions is indicated in (\ref{ordereps}).
The final term in (\ref{singletKeff}) is one of the the Kahler
potential operators (\ref{kahlersix}) which can modify the $H_u$
potential at effective dimension six. With $F_{H_d} \simeq -( \mu
H_u)^{\dagger}$, it gives an $\eta$-independent correction to the
light Higgs mass at order $\epsilon^2$. In the normalization
(\ref{deltaepss}) this correction is
 \beq \epsilon_3^2 = - 4
 \left| { \mu \lambda_S \over M_S} \right|^2 . \label{eps3singlet}
 \eeq
 This contribution is
negative definite, so it tends to decrease the Higgs mass.  In the
supersymmetric limit, it dominates for small $\eta$.  Even for
moderate $\tan \beta=1/\eta $ (of order 3 -- 4), the contributions
we have enumerated are substantial. Including supersymmetry
breaking effects, there are additional contributions, however, to
$\epsilon_3^2$, which can have either sign.

% --------------------------------

\subsection{Triplets}
\label{sec:triplets}

For the mass of the lightest Higgs, the effects of singlets are
suppressed for small $\eta$, because they only couple to the Higgs
combination $H_u H_d$, so it is interesting to consider
triplets~\cite{triplets,mohapatra}.   We will introduce two such fields, $T,
\bar T$, with hypercharge $+2$ and $-2$, respectively.  As in the
case of the singlets, we will take the triplets to be heavy, with
Kahler potential and superpotential
 \beq \int d^4\theta \left( H_u^{\dagger} e^V H_u
 + H_d^{\dagger} e^V H_d + T^{\dagger} e^V T +
 \overline{T}^{\dagger} e^V \overline{T} \right)\eq
  \beq \int d^2 \theta \left( \mu H_u H_d +
 M _T T \overline{T} + \lambda_T T H_u H_u
 +  \lambda_{\overline{T}} \overline{T} H_d H_d \right).
 \eq
Note the couplings to $H_u H_u$ and $H_d H_d$. To leading order
the heavy triplets may be integrated out through the chiral
superfield holomorphic equation of motion, as in the NMSSM;
 \beq
 \bar T = -{\lambda_{\bar T} \over M_T } H_u^2 \qquad ; \qquad T =
 -{\lambda_T \over M_T } H_d^2.\eeq
In the supersymmetric limit this gives a tree-level Kahler and
superpotential
 \beq
 \int d^4\theta \left( H_u^{\dagger} e^V H_u + H_d^{\dagger} e^V
 H_d + {\vert \lambda_T \vert^2 \over M_T ^2}  (H_u^\dagger e^V
 H_u)^2 +  {\vert \lambda_{\bar T} \vert^2 \over M_T ^2}
 (H_d^\dagger e^V H_d)^2 \right)
           \label{tripletKeff}
 \eq
 \beq \int d^2 \theta \left( \mu H_u H_d - {\lambda_T
 \lambda_{\overline{T}} \over M_T } (H_u H_d)^2 \label{tripletWeff}
 \right).
 \eq

{}From (\ref{tripletWeff}) we can read off the coefficient of the
operator of (\ref{genHsuper}):
 \beq
 {\lambda \over M} = -{\lambda_T
 \lambda_{\overline{T}} \over  M_T }  \qquad ;
 \qquad \epsilon_1 = -{\mu^* \lambda_T
 \lambda_{\overline{T}} \over  M_T }.
 \eeq
Supersymmetry-breaking can be described to leading order by a
spurion coupling
 \beq
 \int d^2 \theta~ {\cal Z}{1 \over 2}M_T  T \overline{T} .
 \eeq
After integrating out the triplets with this soft breaking, the
operator (\ref{WZHHHH}) is generated with
 \beq
 \epsilon_2 = {m_{\rm SUSY} \lambda_T \lambda_{\overline{T}}
 \over M_T}. \eeq
As in the singlet model, there are calculable effects of order
$\epsilon^2$. At zeroth order in $\eta$, the leading contributions
come from the term (\ref{ordereps}), the third term in the Kahler
potential (\ref{tripletKeff}) and from the additional spurion
couplings \beq \int d^4 \theta ~{\cal Z} {\cal Z}^\dagger
(T^\dagger e^V T). \eeq Unlike the singlet case, the sign of the
$\eta$-independent order $\epsilon^2_3$ term depends on the
precise value of the soft breaking in this final operator.

% ----------------------

\subsection{Additional Gauge Interactions}

Another approach to generating quartic couplings has been widely
studied \cite{gaugeref,gaugerefb}.  Gauge interactions beyond
those of the Standard Model, broken at a scale $M_V$, can generate
new contributions to the quartic couplings. In the supersymmetric
limit, properly integrating out the massive gauge fields at $M_V$
eliminates the quartic coupling associated with these gauge
interactions, but creates higher order terms in the Kahler
potential, as in (\ref{kahlersix}). These include supersymmetric
terms and supersymmetry violating terms. The former are of order $
\mu^2/\Lambda^2$ ($\Lambda$ is the vev which breaks the gauge
symmetry) and the latter are of order $ m_{\rm SUSY}^2/\Lambda^2$.

We illustrate the basic phenomenon in $U(1)'$ theories. We add to
the MSSM two charged fields, $\phi^{\pm}$, and a neutral field,
$\phi^0$, and assume that some of the MSSM fields, e.g.\ $H_u$,
are charged under $U(1)'$.  The superpotential is
 \beq
\int d^2 \theta \left( \phi_0(\phi_+\phi_- - \Lambda^2) + W_{\rm
MSSM}  \right),
 \eq
where $W_{\rm MSSM}$ is the MSSM superpotential.  The Kahler
potential is
 \beq
\int d^4 \theta \left(  \phi_+ ^\dagger e^{ V'} \phi_+ + \phi_-
^\dagger e^{- V'} \phi_-  +  K_{\rm MSSM} \right),
 \label{Ksdef}
 \eq
where $V'$ is the $U(1)'$ gauge superfield, and $K_{\rm MSSM}$ is
the MSSM Kahler potential with appropriate insertions of $e^{\pm
V'}$.

Consider, first, this model without supersymmetry breaking.
Ignoring covariant derivative terms, the
$\phi_0$ and $\phi_\pm$ equations of motion lead to
 \beq
 \phi_+\phi_- = \Lambda^2 \qquad , \qquad \phi_0=0,
 \label{phipmeom}\eq
and therefore the gauge symmetry is Higgsed.

We want to integrate out the massive fields $\phi_\pm$, $\phi_0$
and $V'$.  We use the unitary gauge $\phi_+=\Lambda$, and then the
equations of motion (\ref{phipmeom}) allow us to set
 \beq  \phi_+  = \phi_-  =\Lambda.
 \label{phipmg}
 \eeq
(Note that in the
equations of motion leading to (\ref{phipmeom}) we neglected the
kinetic terms.  Including them affects only higher derivative terms in
the effective action.) Using (\ref{phipmg}) in the Kahler potential
we find an effective Kahler potential for $V'$
 \beq
 \int d^4 \theta  \left(
 |\Lambda|^2\left( e^{V'} + e^{-{V'}}\right)
  + K_{\rm MSSM} \right)  =   \int d^4 \theta \left(
     |\Lambda|^2 \left(2 + V'^2+ {\cal O}(V'^4)\right)
  + K_{\rm MSSM}
  \right)
  \label{gaugefa}
 \eq
and hence the gauge boson mass is
 \beq
 M_{V'}^2 = 4 g_{V'}^2 \vert \Lambda \vert^2.
 \eeq
It is now straightforward to use (\ref{gaugefa}) and integrate out $V'$.
Its equation of motion is
 $$
 |\Lambda|^2\left( e^{V'} - e^{-{V'}}\right)
  + D_{\rm MSSM}= 2|\Lambda|^2V'\left( 1+{\cal O}(V'^2)\right)
  + D_{\rm MSSM} =0
  $$
  \beq
   D_{\rm MSSM} = {\delta K_{\rm MSSM} \over \delta V'} .
 \label{Veom} \eq
We recognize the $\theta=\bar\theta=0$ component of this equation as the
standard D-term equation.  As above, here we neglect the kinetic term for
$V'$ because we are interested only in terms without covariant
derivatives in the low energy theory.  We can now solve for $V'$
 \beq
 V' =  -{D_{\rm MSSM}(V'=0) \over 2\Lambda^2} + {\cal O}(1/|\Lambda|^4).
 \label{Vvev}
 \eq
Substituting this in (\ref{gaugefa})
with the expansion
 \beq
 K_{\rm MSSM} = K_{\rm MSSM}(V'=0) + V'{\delta K_{\rm MSSM} \over \delta V'}
  (V'=0) + {\cal O} ( V^{'2})
 \eq
we find the effective Kahler potential
 \beq
 \int d^4 \theta
 K_{\rm eff}  = \int d^4 \theta  \left(
 K_{\rm MSSM}(V'=0) - {D_{\rm MSSM}^2(V'=0)\over 4
 |\Lambda|^2} + {\cal O}(1/|\Lambda|^4) \right).
 \label{Keff}
 \eq
We note that unlike the examples of integrating out chiral
superfields where the quartic correction to the Kahler potential
was positive, here it is negative.

It is now easy to repeat this calculation with explicit
supersymmetry breaking.  For simplicity, we add equal masses for
$\phi_\pm$; i.e.\ we replace (\ref{Ksdef}) with
 \beq
  \int d^4 \theta \left(
 (1-m_{\rm SUSY}^2 \theta^4)\left( \phi_+ ^\dagger e^{ V'} \phi_+
 + \phi_- ^\dagger e^{- V'} \phi_-\right)+  D_{\rm MSSM} \right).
 \label{Ksdefs} \eq
Then, (\ref{Keff})  becomes
 \beq
 % \int d^4 \theta K_{\rm eff} =
  \int d^4 \theta \left(
  K_{\rm MSSM}(V'=0) - (1+m_{\rm SUSY}^2 \theta^4)
 {D_{\rm MSSM}^2(V'=0)\over 4 |\Lambda|^2} + {\cal O}(1/|\Lambda|^4)
 \right).
 \label{Keffs} \eq

We are interested in the contributions which depend on $H_{u,d}$.
{}From (\ref{Keffs}) we find
 \beq
  - \int d^4 \theta \left((1+m_{\rm SUSY}^2 \theta^4)
 {1\over 4 |\Lambda|^2} \left(q_{H_u}(H_u^\dagger e^V H_u)
 + q_{H_d} (H_d^\dagger e^V
 H_d) \right)^2
  \right)
  \label{kahcov}
 \eq
where $q_{H_{u,d}}$ are the $U(1)'$ charges of $H_{u,d}$.

We note that in this model there are no superpotential corrections
and therefore $\epsilon_1=\epsilon_2=0$.  We recognize in the
Kahler potential correction (\ref{kahcov}) some of the operators
in (\ref{kahlersix}) with specific relations between their
coefficients $\xi_i$. In particular, setting $M = 2 |\Lambda|$, we
have $\xi_2=-2 q_{H_u}q_{H_d}$, $\xi_3= -q_{H_u}^2$, and $\xi_5=-
q_{H_u}^2$ and all other $\xi_i$ vanish. Using these $\xi_i$ it is
straightforward to compute $\epsilon_3$ and the shift in $m_h^2$.
In general, there are additional supersymmetry breaking parameters,
which will contribute to the $\xi_i$'s.

There are a number of phenomenological constraints which must be
satisfied. Typically it is necessary that $M_V > 1$ TeV; moreover,
the gauge coupling constants might be expected to be small (the
usual $U(1)$ coupling in the Standard model is about 1/3).  In
such circumstances, the scale $m_{\rm SUSY}$ must be of order
$500$ GeV or larger if the corrections to the quartic coupling are
to account for the light Higgs mass.  If $\mu \sim 200$ GeV, then
the effects of $\xi_2$ and $\xi_3$ are negligible, and
$\delta_{\epsilon^2}m_h^2 \simeq q_{H_u}^2 m_{\rm
SUSY}^2v^2/|\Lambda|^2$.

While $M_V$ must be rather large, and as a result, the
supersymmetry breaking scale in this sector cannot be too small,
there are naturalness upper bounds on these scales as well.  It is
important that loop corrections to scalar masses of MSSM fields
which carry the additional gauge quantum numbers not be too large;
this, combined with the requirement that the quartic correction be
substantial, constrains the scale $M_V$.  It is possible that
$m_{\rm SUSY}$ is comparable to $M_V$, and that an operator
analysis is not appropriate. Simple models of this type can be
obtained by taking no superpotential for $\phi^{\pm}$, and
different soft breaking masses for each field.

An interesting model of this kind can be motivated from familiar
compactifications of the heterotic string theory.  In $E_6$, there
are two $U(1)$'s beyond those of the Standard Model.  It is
possible to break one of these, say $B-L$ at very high energies.
If one assumes the remaining $U(1)$ is broken at a scale of, say,
a few TeV, and that the scale of supersymmetry breaking in the
vector multiplet is comparable, a conventional unification
calculation leads to a small value of the $U(1)$ coupling, but can
readily give a light Higgs mass of order $125$ GeV or so.
Finally, theories with non-Abelian symmetries can give much larger
Higgs mass \cite{gaugerefb}.

% ------------------------------------------------------

\section{Conclusions}

The current limit on the Higgs mass may suggest that if nature is
supersymmetric, the underlying model contains degrees of freedom
beyond those of the MSSM. If this physics lies at scales a bit above
that of the MSSM degrees of freedom, its effects are naturally
organized within an effective Lagrangian. In the MSSM Higgs sector
there are only two operators at effective dimension five. These
operators can have important effects on the scalar Higgs masses. For
the light Higgs mass at very large $\tan \beta$, it is necessary to
consider operators of effective dimension six; there are potentially
five such operators, though many are negligible in simple modes.

It is conceivable that the puzzle of the Higgs mass is resolved by
very massive stops, or by large stop mixing.  It seems more
plausible that, if the hypothesis of low energy supersymmetry is
correct, stop squarks with more modest masses and mixings will be
discovered. Once the stop masses and mixing are known we can
calculate the radiatively corrected Higgs masses and compare them
with the measured values.  If a discrepancy with the simple MSSM
predictions is found, then the BMSSM operators can correct the
masses. At leading order in $\epsilon$, the three masses $m_h$,
$m_H$ and $m_{H^\pm}$ are corrected only by two real numbers
$\epsilon_{1r}$ and $\epsilon_{2r}$.  Therefore, the measured
values over-constrain these two unknowns.  In particular, for
small $\eta$ the masses of $H$ and $H^\pm$ are corrected only by
one real number $\epsilon_{2r}$, and therefore the BMSSM relation
$2m_{H^{\pm}}^2 = m_H^2 + m_A^2 + 2 m_W^2 + {\cal O}( \eta^2, \eta
\epsilon, \epsilon^2 )$ (up to radiative corrections) can be
tested.

The new operators of the BMSSM can have interesting effects not
only on the Higgs mass spectrum but on decays as well. For
example, the Higgs-Higgsino interactions (\ref{higgshiggsino})
contribute to decay of heavy Higgses to charginos and neutralinos
$H,A \rightarrow \chi_i^+ \chi_j^-, \chi_i^0 \chi_j^0$ and
$H^{\pm} \rightarrow \chi_i^{\pm} \chi_j^0$ for both $i=j$ and $i
\neq j$. These interactions also contribute to neutralino and
chargino decays which involve Higgs bosons, $\chi_i^0 \rightarrow
h \chi_j^0, H \chi_j^0, A \chi_j^0, H^{\pm} \chi_j^{\mp}$ and
$\chi_i^{\pm} \rightarrow H^{\pm} \chi_j^0, h \chi_j^{\pm}, H
\chi_j^{\pm}, A \chi_j^{\pm}$. In regions of parameter space where
the initial and final state charginos or neutralinos are
Higgsino-like the operator (\ref{higgshiggsino}) can significantly
modify these branching ratios compared with the tree-level MSSM.
The interactions (\ref{higgshiggsino}) can also lead to three body
decays such as $H,A \rightarrow h \chi_i^+ \chi_j^-, h \chi_i^0
\chi_j^0$ and $\chi_i^0 \rightarrow hh \chi_j^0$. CP violation in
the effective dimension five operators also opens interesting
decay modes which are absent in the tree-level MSSM, such as $A
\rightarrow hh$ and $A \rightarrow WW,ZZ$ through mixing.

Future precision fits to chargino and neutralino masses, mixings,
and couplings in both production and decay would also be affected.
However, it is important to stress that at effective dimension
five all these observables depend only on $\epsilon_1$, and thus
represent a highly over-constrained parameter space for the
precision fits. So restricting to the BMSSM effective dimension
five parameter space could still allow for non-trivial tests of
both supersymmetry and the extension of the MSSM.

The BMSSM operators can also have interesting implications for
cosmological SUSY signatures. Details of electroweak baryongenesis
could be affected in a number of ways.\footnote{We thank Ann
Nelson for pointing out this possibility.} First, the phase
transition could be somewhat more strongly first order than in the
MSSM as currently constrained, since both of the stop squarks
could be fairly light if the BMSSM operators contribute
significantly to the zero temperature Higgs mass. Second, the
quartic Higgs couplings can provide an interesting source of CP
violation within the bubble wall. Finally, the Higgs-Higgsino
interactions with CP violating phases could contribute to an axial
current of Higgsinos scattered from the bubble wall (which is
ultimately processed into a baryon asymmetry by sphalerons).

The relic abundance of neutralino dark matter in the well know
gaugino-Higgsino region of parameter space is also modified in the
BMSSM. In this region of the MSSM the observed relic dark matter
abundance is obtained for the $\mu$-parameter and the Bino or Wino mass,
$m_1$ or $m_2$, very close in value, resulting in sizeable Bino-Higgsino
or Wino-Higgsino
mixing. Details of this mixing as well as couplings to final
states with or through Higgs bosons are sensitive to the dimension
five Higgs-Higgsino interactions. In particular, CP violation in
these interactions would open up $S$-wave annihilation channels
through light $s$-channel Higgs bosons, and could quantitatively
shift the region of parameter space which results in the proper
relic abundance.

% ----------------

 \bigskip
  \centerline{\bf Acknowledgments}
We would like to thank Nima Arkani-Hamed, Ann Nelson and Lisa Randall
for useful discussions. M.D. and S.T. thank the Institute for Advanced Study
for its hospitality.  The work of M.D. was supported by the US Department
of Energy. The work of N.S. was supported in part by DOE grant
DE-FG02-90ER40542. The work of S.T. was supported in part by DOE
grant DE-FG02-96ER40959.

% ------------------------------------------------------------

\appendix

\section{Two Doublet Higgs Sector Symmetries }

The most general $SU(2)_L \times U(1)_Y$ invariant renormalizable
scalar potential for two Higgs doublets, $H_u$ and $H_d$, with
hypercharge $Y= \pm 1$, is given in standard notation by
\cite{rghh}
$$
V = \widetilde{m}_{H_u}^2 H_u^{\dagger}H_u + \widetilde{m}_{H_d}^2 H_d^{\dagger}H_d
     - \left( m_{ud}^2 H_u H_d + ~{\rm h.c.} \right) %~~~~~~~~~~~~~~~~~~~~~
+ {1 \over 2} \lambda_1 (H_u^{\dagger} H_u)^2
$$
$$
+ {1 \over 2} \lambda_2 (H_d^{\dagger} H_d)^2
+ \lambda_3 (H_u^{\dagger} H_u)(H_d^{\dagger} H_d)
+ \lambda_4 (H_u^{\dagger} H_d)(H_d^{\dagger} H_u)
$$
\beq
~~~~+ \left[ {1 \over 2} \lambda_5 (H_u H_d)^2 +
        \left( \lambda_6 (H_u^{\dagger} H_u)
       + \lambda_7 (H_d^{\dagger} H_d) \right) H_u H_d
      ~+~{\rm h.c.} \right]
 \label{Vmostgeneral}
\eq
The Higgs potential (\ref{VMSSMtree}) of the renormalizable MSSM,
along with the effective dimension five interactions
(\ref{vepsilon1}) and (\ref{genAcorrection}) correspond to the
potential (\ref{Vmostgeneral}) with
$$
\lambda_1 = \lambda_2 = {1 \over 4} (g^{\prime 2} + g^2)
$$
$$
\lambda_3 = {1 \over 4} (g^2 - g^{\prime 2}) ~~~~~~
$$
$$
\lambda_4 = - {1 \over 2} g^2 ~~~~~~~~~~~~~
$$
$$
\lambda_5= 2 \epsilon_2 ~~~~~~~~~~~~~~~~
$$
\beq
\lambda_6= \lambda_7 = 2 \epsilon_1 ~~~~~~~~~
\eq

In order to classify patterns within the physical Higgs mass
spectrum and interactions it is useful to determine how the Higgs
sector couplings transform under background global symmetries. The
largest possible symmetry of the scalar potential
(\ref{Vmostgeneral}) is an $SO(8)$ under which the eight real
components of $H_u$ and $H_d$ transform in the ${\bf 8}_v$
representation. An interesting subgroup of this maximal symmetry
is $SU(2)_{L_u} \times SU(2)_{R_u} \times SU(2)_{L_d} \times
SU(2)_{R_d} \subset SO(8)$ under which the Higgs fields transform
as bi-doublets
$$
\begin{array}{ccccc}
    & SU(2)_{L_u} & SU(2)_{R_u} & SU(2)_{L_d} & SU(2)_{R_d} \\
    &  &  &  &  \\
H_u &  {\bf 2}  & {\bf 2} & {\bf 1}  & {\bf 1}  \\
H_d &  {\bf 1}  & {\bf 1} & {\bf 2}  & {\bf 2}    \\
\end{array}
$$
%A convenient representation of the Higgs fields under this decomposition
%is
%\beq
%{\cal H}_{u~a}^{~\alpha} = \left(
%\begin{array}{cc}
%H_u^{0*} & H_u^+ \\
%H_u^- &  H_u^0
%\end{array} \right)
%~~~~~~~~
%{\cal H}_{d~a}^{~\alpha} = \left(
%\begin{array}{cc}
%H_d^{0} & H_d^+ \\
%H_d^- &  H_d^{0*}
%\end{array} \right)
%\label{bidoublets}
%\eq
%where $\alpha$ and $a$ are $SU(2)_L \times SU(2)_R$ doublet indices.
The $SU(2)_L \times U(1)_Y$
gauge symmetry is embedded in the diagonal subgroups of
this decomposition as
$$
~~~~~~~~~~~~
SU(2)_L \subset SU(2)_{L_u} \times SU(2)_{L_d}
$$
\beq
U(1)_Y \subset SU(2)_R \subset SU(2)_{R_u} \times SU(2)_{R_d}.
\label{gaugebed}
\eq

Expectation values for the Higgs fields spontaneously break each
$SU(2)_L \times SU(2)_R$ to a diagonal
custodial subgroup
$$
\langle H_u \rangle ~~~:~~~~SU(2)_{L_u} \times SU(2)_{R_u}
   \rightarrow SU(2)_{C_u}
$$
\beq
\langle H_d \rangle ~~~:~~~~SU(2)_{L_d} \times SU(2)_{R_d}
   \rightarrow SU(2)_{C_d}
   \label{breakings}
 \eq
The diagonal subgroup of these custodial symmetries
 \beq SU(2)_C
\subset SU(2)_{C_u} \times SU(2)_{C_u} \label{custodial} \eq
provides a useful symmetry for understanding features of the Higgs
spectrum. It reduces to the usual $SU(2)_C$ custodial symmetry of
the one Higgs doublet model in the limit in which the second Higgs
doublet is decoupled. Since the most general expectation values
(\ref{breakings}) leave $SU(2)_C$ unbroken, any violations arise
from the interactions.

If the potential preserves the $SU(2)_C$ symmetry, then the three
eaten Goldstone modes are in an $SU(2)_C$ triplet, the two massive
Higgses, $h$ and $H$ are $SU(2)_C$ singlets, and there is a
massive triplet $H^+,A,H^-$. The couplings of the Higgs potential
(\ref{Vmostgeneral}) transform as the components with vanishing
$U(1)_Y \subset SU(2)_C$ representations
 \beq
\begin{array}{cc}
           & SU(2)_C \\
           & \\
\widetilde m_{H_{u,d}}^2 ,\ \
{\rm Re}(m_{ud}^2 e^{i \varphi}),\
\ \lambda_{1,2,3},\ \
\lambda_4 + {\rm Re}(\lambda_5 e^{2 i \varphi}),\ \
{\rm Re}(\lambda_{6,7} e^{ i \varphi})  &  {\bf 1} \\
{\rm Im}(m_{ud}^2 e^{i \varphi}),\ \  {\rm Im}(\lambda_5 e^{2 i \varphi})
 ,\ \ {\rm Im}(\lambda_{6,7} e^{ i \varphi})    & {\bf 3} \\
\lambda_4 - {\rm Re}(\lambda_5 e^{2 i \varphi})  &  {\bf 5} \\
\end{array}
 \eq
where in a general basis $\varphi$ is the phase of the expectation
value
 \beq
\varphi = {\rm Arg}\langle H_u H_d \rangle.
\eq

If CP symmetry is unbroken, either explicitly or spontaneously,
there is a basis in which $\varphi=0$ and the imaginary components of
all potential couplings vanish. In this basis the $SU(2)_C$
transformation properties of the Higgs potential couplings reduce
to those under the diagonal $SU(2)_R$ in (\ref{gaugebed}), since
the phase $\varphi$ vanishes and all couplings are $SU(2)_L$
invariant. In addition, the component field $A$ is a mass
eigenstate if CP is unbroken. Since $H^+, A, H^-$ transform as a
${\bf 3}$ of $SU(2)_C$ the mass squared terms for these fields
transform as the symmetric representations ${\bf 1} + {\bf 5}
\subset {\bf 3} \times {\bf 3}$ of $SU(2)_C$
 \beq
V = m_{\bf 1}^2 \left( {1 \over 2} A^2 + H^+ H^- \right) +
m_{\bf 5}^2 \left( {1 \over 2} A^2 - H^+ H^- \right)
\  + \cdots
 \eq
In the CP conserving limit, $\lambda_4 - \lambda_5$ is the only
combination of renormalizable Higgs potential couplings which
transforms under ${\bf 5}$ of $SU(2)_C$ and leads to splittings
between $H^{\pm}$ and $A$
 \beq m_{H^{\pm}}^2 = m_A^2 + v^2
(\lambda_5 - \lambda_4). \eq

Another useful global symmetry which provides selection rules is
the well known $U(1)_{PQ}$ symmetry under which both $H_u$ and
$H_d$ have the same charge.  This symmetry is generated by the
Cartan subalgebra element orthogonal to the gauged $U(1)_Y$
 \beq
 ~~~~~~~~~~~~ U(1)_{PQ} \subset SU(2)_{R_u} \times SU(2)_{R_d}.
 \eq
If the potential preserves this $U(1)_{PQ}$ symmetry, the massive
fields $H^\pm$ and $H\pm iA$  are charged under it and therefore,
$H$ and $A$ are degenerate.

The terms in the more general potential are classified by the
$U(1)_{PQ}$ charge as follows
 \beq
\begin{array}{cc}
           & U(1)_{PQ} \\
           & \\
\widetilde m_{H_{u,d}}^2,\ \  \lambda_{1,2,3,4} \ \       & 0 \\
m_{ud}^2 , \ \ \lambda_{6,7}  &  -2  \\
\lambda_5  & -4   \\
\end{array}
 \eq
One application of this symmetry is the limit $m_{ud}^2 ,\
\lambda_{6,7} \to 0 $ with $\widetilde{m}_{H_u}^2  <0$ and
$\widetilde{m}_{H_d}^2 > 0$ for which only the up-type Higgs gains
an expectation value.  This is the large $\tan \beta$ limit of the
(B)MSSM with $m_A$, or equivalently $\widetilde{m}_{H_u}^2 $ and
$\widetilde{m}_{H_d}^2 $, held fixed.

% ---------------------------------------------------------------------

\end{document}